\documentclass[a4paper]{article}

\usepackage{graphicx}
\usepackage[cm]{fullpage}
\usepackage{multicol}
\usepackage{wrapfig}
\usepackage{xcolor}
\usepackage{colortbl}
\usepackage[overlay]{textpos}
\usepackage{ragged2e}

\begin{document}

\begin{center}
\Huge
\textbf{Towards the computational experiment}
\end{center}

\textit{This article was originally published in Arkhimedes, the magazine of the Finnish
Mathematical and Physical Societies, which own its copyright. It is reproduced here,
with permission, to make it available to the international public. Citation to this
work should refer to the original bibliographical source material: M.~A. Caro.
Arkhimedes \textbf{3}, 21 (2018).}

\begin{textblock}{0.25}(0.05,10.5)
\begin{center}
\begin{tabular}{|>{\columncolor[RGB]{255,230,210}}p{5.2cm}|}
\textit{Dr. \textbf{Miguel Caro} is an Academy of Finland postdoctoral researcher at
Aalto University. For more information on computational simulation of materials, he can
be contacted on miguel.caro@aalto.fi or mcaroba@gmail.com.}
\end{tabular}
\end{center}
\end{textblock}

\begin{multicols}{3}
Imagine it's the year 2266 and the crew of the \textbf{\itshape Rocinante}, a salvaged
Martian military spaceship, has been sent on a diplomatic mission to a newly discovered
planet located in a distant point of the galaxy. The newly arrived human inhabitants soon
discover that some of the local creatures produce chemical compounds that are toxic to
Earth-based life. Since real laboratories are not available in this remote planet and help
from Earth would take too long to arrive, the \textbf{\itshape Rocinante}'s crew and local
scientists collect field data that is then relayed using antennas to state-of-the-art
supercomputing facilities back in the solar system. The large computers back home are able
to run a series of computer simulations to understand the alien chemistry and help find a
cure to the new diseases threatening their lives.
\\
\indent
These fictional events take place in James S.A. Corey's \textbf{\itshape Cibola Burn}
(2014), a novel which is part of the Expanse series, now turned into a successful TV show.
The events it describes, and the idea that real experiments can be effectively replaced by
computational experiments, sound very much like what they are: science fiction. Or, do
they? Much of the general public, and perhaps quite a few (experimental) scientists, would
be surprised to learn that computational experimentation and, for that matter,
\textit{accurate} computational experimentation, is not as much fiction as it is a reality
-- with severe limitations.
For quite some time now, computational physicists and chemists
have been able to accurately model molecular
systems made up of a few atoms, such as small molecules. With the introduction of density
functional theory\linebreak

\vspace{8em}

\noindent
(DFT) in the 1960s (an
achievement recognized with the Nobel prize awarded to Walter Kohn in 1998) and the
development of better algorithms, faster computers, and parallel computer architectures,
in recent years running accurate simulations of systems with a couple hundreds of atoms has
become feasible. For instance, we can get a reasonably accurate description of the structure
of small organic molecules in aqueous solution, such as the glutamate molecule surrounded by
water shown in Fig.~1.

\begin{center}
\hrule
\vspace{0.2em}
\includegraphics[width=\linewidth]{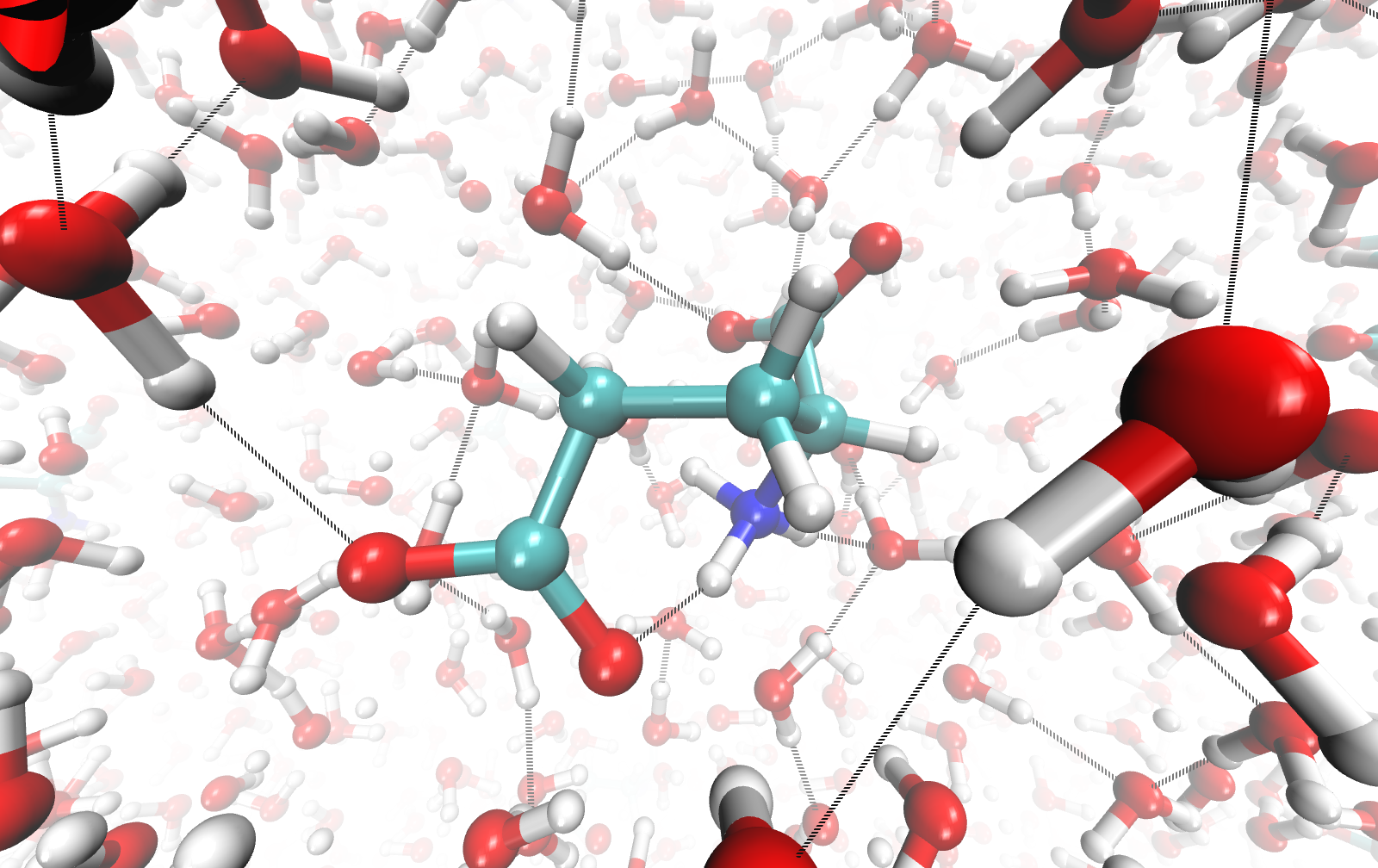}
\\
\justify{
\small Fig.~1: Water forms complicated networks around solvated compounds, like this glutamate
molecule (a neurotransmitter). These networks  evolve quickly with time, with characteristic
time scales in the picosecond (10$^{-12}$ seconds) range.
}
\hrule
\end{center}

The dynamical water solvation effects can be studied by
solving Newton's equations of motion for the nuclei, which are treated as classical particles,
while the electrons are modeled using DFT. This approximation is known as the Born-Oppenheimer
approximation, and this type of simulation is known as \textit{ab initio} molecular dynamics (MD). For
a system of the size shown in Fig.~1, with about 200 atoms, a 10 nanosecond ($10^{-8}$
seconds) simulation can take approximately one million CPU hours to run. We can run these
simulations only on large parallel computers, that is, unless we found a scientist willing
to spend 114 years running this calculation on a single-core computer, assuming that he or
she would live that long (and, even more unlikely than finding such long-lived scientist,
assuming that we could convince a funding agency to give us the money to keep this project
running for 114 years!).
\\
\indent More complicated problems, such as those arising in biology and medicine, cannot currently
be accurately solved using computer simulation of the atomic interactions at the DFT level.
This is due to the long time scales and system sizes involved. For instance, a typical
medium-sized protein found in the human body contains about 5000 atoms, and molecular processes
that are of interest to biochemists, such as protein folding, take about 10 microseconds
($10^{-5}$  seconds) to occur. So, assuming linear scaling with system size and
time,\footnote{The assumption of linear scaling with time is realistic; the assumption or
linear scaling with system size is... optimistic.} and extrapolating from our small system
from before, it would take 25
\textit{billion} CPU hours to run this simulation using \textit{ab initio} MD (and we have
not even considered the water molecules that make up the solvation shell of the protein!).
For reference, the Finnish national supercomputing center, CSC, provides about 500 million CPU
hours per year for the whole computational science community of Finland (as of 2018). In other
words, a simulation of this kind would use up all the scientific computing resources available
for the whole country over several decades. In practice, systems of these sizes are currently
modeled using empirical classical potentials, which treat the interactions between atoms in an
effective way, using approximations such as harmonic bonds and bond angles, where pairs and
triplets of atoms are allowed to move as if they were connected by springs (Fig.~2).
These simulations are many orders of magnitude cheaper than DFT, to the point that with classical
potentials, one can run the CPU-hungry protein folding simulation on a laptop in the morning and
have it ready before lunch time. Unfortunately, this computational efficiency comes at a cost
-- the loss of accuracy.

Classical potentials are good for resolving harmonic (``spring-like'')\linebreak


\begin{textblock}{9.}(4.7,9.3)
\begin{center}
\hrule
\vspace{0.2em}
\includegraphics[width=10cm]{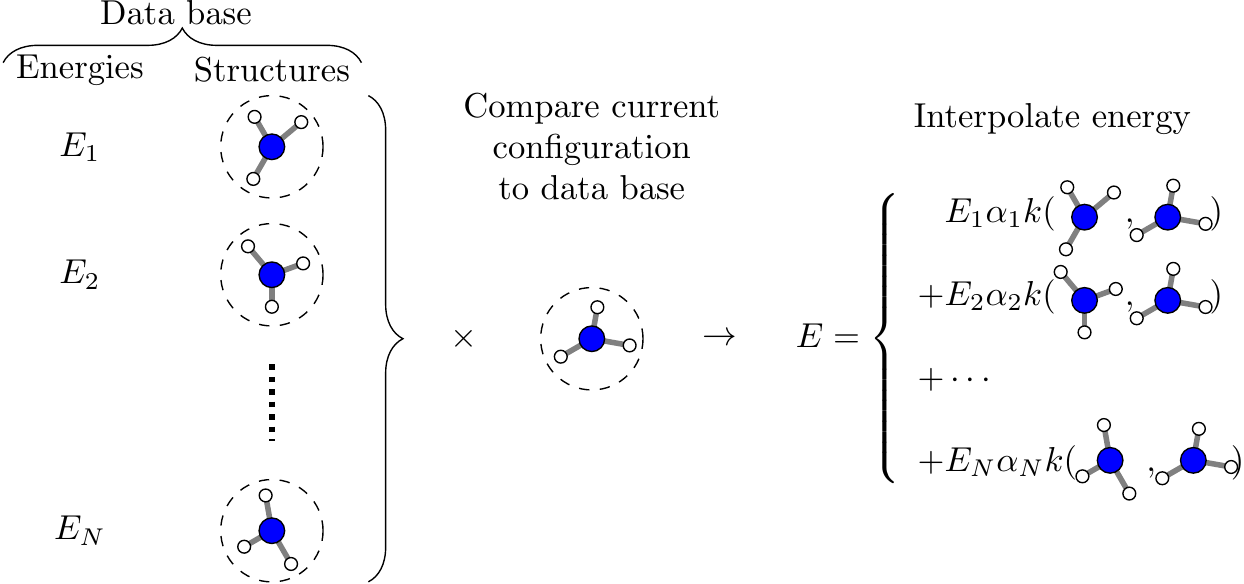}
\\
\justify{
\small Fig.~3: GAP potentials compare the current atomic structure, ammonia in this example, to all
the structures in the data base, for which energies have been precomputed. The predicted  energy
$E$ is given as the sum of the energies in the data base, $E_1, E_2, \dots, E_N$, times a fitting
coefficient, $\alpha$, times the kernel (similarity) between the present structure and those in the
data base, $k$. The fitting coefficients are also precomputed when training the model.
}
\end{center}
\end{textblock}

\begin{center}
\includegraphics[width=0.8\linewidth]{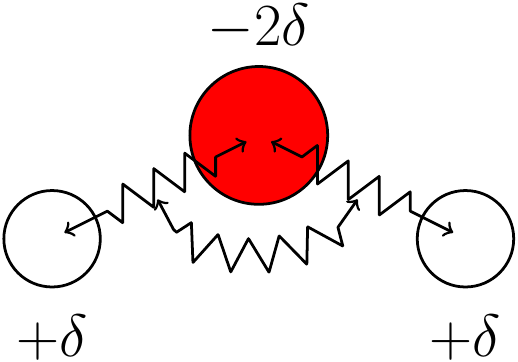}
\\
\justify{
\small Fig.~2: Classical potentials often  model atomic bonds as if they  were springs (i.e.,
stronger bonds are modeled with stiffer springs). Complicated electrostatic interactions are
approximated with partial charges. This example shows a water molecule.
}
\vspace{0.2em}
\hrule
\end{center}

atomic vibrations. Even
electrostatic interactions can be included, thus improving the model, in a limited way by
assuming partial charges. For example, the oxygen atom in a water molecule is partially negatively
charged, since it tends to attract electrons, while the hydrogens are partially positively
charged, since they tend to give up electrons (Fig.~2). What classical potentials are not
good at is describing the change in these effective interactions as the atomic environment changes.
This is bad news, since during a molecular dynamics simulation the atomic environments are
constantly changing. How do the spring constants or the partial charges in a water molecule change
as the water molecule is deformed? Or when it is surrounded by another molecule? Or when the
molecule is \textit{broken}? Classical potentials are highly constrained by their fixed functional form,
they are bad at describing molecules away from equilibrium, and they are \textit{extremely bad} at
describing the formation and breaking of chemical bonds. For instance, a classical potential
will fail at reproducing the combustion of two H$_2$ molecules and one O$_2$ molecule to give two
H$_2$O molecules, since the H--H and O--O bond breakings and subsequent O--H bond formation are
incompatible with the form itself of the potential. This lack of flexibility also hinders classical
potentials. How can we combine the accuracy and flexibility of DFT with the computational efficiency
of classical potentials? How can we carry out accurate simulations for large systems in a reasonable
amount of time? The most promising solution comes not from the physics or chemistry communities, but
from computer science and data analysis: machine learning.
\\
\indent The fundamental tenet of machine learning applied to modeling atomic interactions is that once we
have run a DFT calculation of several interacting atoms, which has cost us much of our precious CPU
time, why do it again? The main idea is to create a data base of accurate DFT calculations for
small systems that cover the most representative set of ``typical'' atomic structures. The
machine-learning algorithm ``learns'' accurate DFT energies for these structures. Then, to predict
the energy of a new structure, the algorithm ``compares'' it to all the structures in the data base
and interpolates the energy for the new structure, based on all the previously-observed energies
stored in the data base, without the need to run the DFT calculation. This procedure is schematically
depicted in Fig.~3. The measure of similarity between a given atomic ``environment'' and a
previously observed environment in the data base, used to compare them, is known as the kernel. It
varies between 1 (the structures are identical) and 0 (the structures are totally different). Some
of these ideas were pioneered by Dr. Albert Bart\'ok and Prof. G\'abor Cs\'anyi at the University
of Cambridge [1,2]. Cs\'anyi's group has been hard at work on 1) the development of the underlying
theoretical framework, known as Gaussian approximation potential (GAP),\linebreak

\vspace{21em}

\noindent
2) how to create good data
bases to optimally train these GAP models, and 3) how to build good kernels from structural atomic
information. While these machine-learning models are more CPU intensive than classical potentials,
they allow us to run atomistic simulations at almost the same level of accuracy as DFT at roughly
1/1000 of the computational cost (and future code and algorithm development is certain to improve
this figure). But, most importantly, this reduced CPU cost means that we can now do simulations
that were previously out of reach and look into unknown physical and chemical properties of materials.
\\
\indent
Recently at Aalto University, we teamed up with Cs\'anyi and Dr. Volker Deringer (who has developed
a GAP for elemental carbon [3]) to study and understand the reason why so-called tetrahedral
amorphous carbon (ta-C) has mechanical properties similar to those of diamond [4]. Unlike diamond,
ta-C can be made inexpensively in the lab and can be used to protectively coat other materials. It
is used, for instance, to minimize usage wear of titanium hip implants due to friction. With the
GAP approach, we were able to simulate the deposition of energetic carbon atoms onto a preexisting
substrate with system sizes (thousands of atoms) and time scales (one picosecond [10$^{-12}$~s]
per impact) which allowed us to very closely reproduce the experimental conditions. Since we had
direct access to the trajectory of the atoms, we could directly observe how the atoms in the
grow-\linebreak

\vspace{21em}

\newpage

\begin{minipage}{0.5\linewidth}
\vspace{21em}
\end{minipage}

\noindent ing ta-C films developed tetrahedral (four-fold) coordination, just like they have in diamond.
Most interestingly, the growth mechanism that we could infer with these simulations (known as
``peening'', whereby atom packing occurs away from the impact site) was in direct contradiction
with the explanation proposed by experimentalists based on indirect measurements (known as
``subplantation'', whereby\linebreak

\begin{minipage}{0.5\linewidth}
\vspace{20em}
\end{minipage}

\noindent atom packing happens at and around the impact site), as shown in
Fig.~4. In other words, thanks to the power of machine learning, we could run a
computational experiment that gave us information inaccessible from the ``real'' experiment. Now,
our work focuses, among other things, on how to improve these models by ``augmenting'' structural
kernels with electronic structure information. This will allow us to further decrease the error of
the machine-learning interpolation, thus getting closer to the DFT level of accuracy.
\\
\indent Machine learning is certainly not a substitute for DFT (or higher-level quantum chemistry methods),
since a machine-learning interatomic potential still hinges on the need for DFT to generate the
training data. However, machine learning offers a way to do ``cheap DFT'' and, borrowing from the
TV commercial language, to ``\textbf{get more science out of your CPU time}''. The promise of a computational
experiment that can reliably replace a real experiment is not science fiction anymore. The
computational experiment is not the future, \textit{it is the present}.

\vspace{1em}

\hrule

\vspace{1em}

{\scriptsize
\noindent [1] A.P. Bart\'ok, M.C. Payne, R. Kondor, and G. Cs\'anyi. Phys. Rev. Lett. \textbf{104}, 136403 (2010).

\noindent [2] A.P. Bart\'ok, R. Kondor, and G. Cs\'anyi. Phys. Rev. B \textbf{87}, 184115 (2013).

\noindent [3] V.L. Deringer and G. Cs\'anyi. Phys. Rev. B \textbf{95}, 094203 (2017).

\noindent [4] M.A. Caro, V.L. Deringer, J. Koskinen, T. Laurila, and G. Cs\'anyi. Phys. Rev. Lett. \textbf{120}, 166101 (2018).

\vspace{1em}

\hrule

\vfill
}

\begin{textblock}{9.}(-9.3,-6.65)
\begin{center}
\includegraphics[width=10cm]{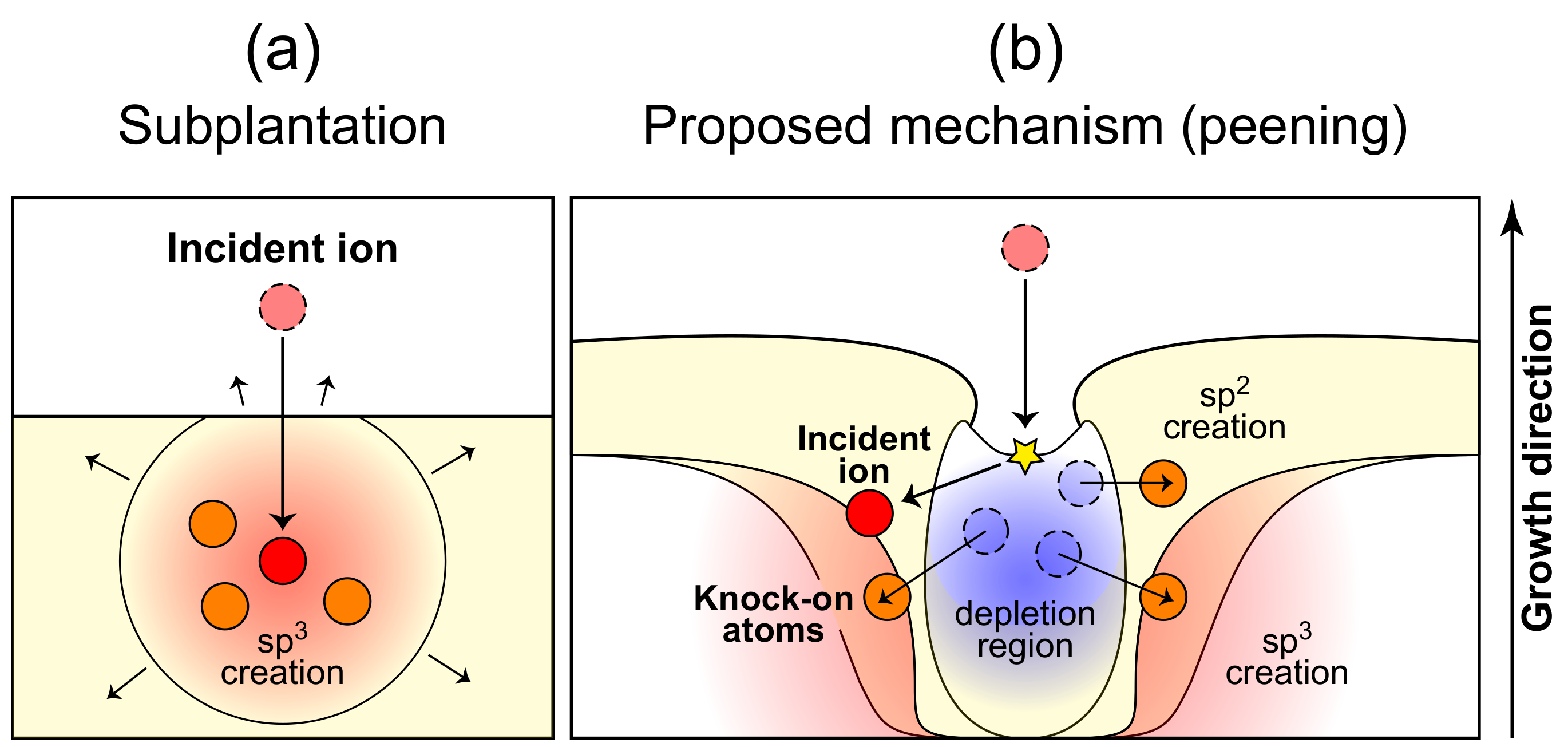}
\\
\justify{
\small Fig.~4: The previously assumed growth mechanism in amorphous carbon, subplantation, was disproven by the
GAP computer simulation. Based on these results, we proposed that the peening mechanism is the predominant
process whereby tetrahedral amorphous carbon acquires its ``diamond-likeness''. \textit{Reprinted
figure with permission from [4]. Copyright (2018) by the American Physical Society.}
}
\vspace{0.2em}
\hrule
\end{center}
\end{textblock}

\end{multicols}

\end{document}